\documentclass[journal]{IEEEtran}
\IEEEoverridecommandlockouts

\usepackage{url, authblk}
\usepackage{graphicx}
\usepackage{amsfonts}
\usepackage{color}
\usepackage{stfloats}
\usepackage{color,soul}
\usepackage{booktabs}

\usepackage{amsmath}
\usepackage{graphicx} 
\usepackage{multirow}
\usepackage{hhline}
\usepackage{url}
\usepackage{float}
\usepackage{enumitem}
\usepackage{cite}
\usepackage{multicol}
\usepackage{color, colortbl}
\usepackage{blindtext}
\usepackage{rotating}
\usepackage{array}
\usepackage{tikz}
\usepackage{pifont}
\usepackage{glossaries}
\usepackage{url}
\usepackage{tabularx}
\usepackage{booktabs}
\usepackage{threeparttable}
\usepackage{subcaption}
\usepackage{soul}

\usepackage[top=0.765in, bottom=1.05in, left=0.625in, right=0.625in]{geometry}
\def\BibTeX{{\rm B\kern-.05em{\sc i\kern-.025em b}\kern-.08em
    T\kern-.1667em\lower.7ex\hbox{E}\kern-.125emX}}

\newcolumntype{P}[1]{>{\centering\arraybackslash}p{#1}}

\newcommand{\veriful}{{{\textsc{VeriFUL}}}}

\newcommand{\uparrowc}{\textcolor{blue}{\boldmath$\uparrow$}}
\newcommand{\downarrowc}{\textcolor{red}{\boldmath$\downarrow$}}
\def\BibTeX{{\rm B\kern-.05em{\sc i\kern-.025em b}\kern-.08em
    T\kern-.1667em\lower.7ex\hbox{E}\kern-.125emX}}
\begin{document}

\title{Towards Verifiable Federated Unlearning: Framework, Challenges, and The Road Ahead}

\author{
\IEEEauthorblockN{Thanh Linh Nguyen\IEEEauthorrefmark{5}\textsuperscript{*},
Marcela Tuler de Oliveira\IEEEauthorrefmark{2},
An Braeken\IEEEauthorrefmark{3},
Aaron Yi Ding\IEEEauthorrefmark{2},
Quoc\mbox{-}Viet Pham\IEEEauthorrefmark{5}}

\IEEEauthorblockA{\IEEEauthorrefmark{5}Trinity College Dublin, Dublin, Ireland}

\IEEEauthorblockA{\IEEEauthorrefmark{2}Delft University of Technology, Delft, The Netherlands}

\IEEEauthorblockA{\IEEEauthorrefmark{3}Vrije Universiteit Brussel, Brussels, Belgium}

\IEEEauthorblockA{Emails: 
\IEEEauthorrefmark{5}{\{tnguyen3, viet.pham\}@tcd.ie}, 
\IEEEauthorrefmark{2}{\{M.TulerdeOliveiraa, Aaron.Ding\}@tudelft.nl}, 
\IEEEauthorrefmark{3}{an.braeken@vub.be}}

\thanks{\textsuperscript{*}Part of this work was completed at TU Delft guided by Aaron Yi Ding.}
}



\maketitle
\begin{abstract}
Federated unlearning (FUL) enables removing the data influence from the model trained across distributed clients, upholding the right to be forgotten as mandated by privacy regulations. FUL facilitates a value exchange where clients gain privacy-preserving control over their data contributions, while service providers leverage decentralized computing and data freshness. However, this entire proposition is undermined because clients have no reliable way to verify that their data influence has been provably removed, as current metrics and simple notifications offer insufficient assurance. We envision unlearning verification becoming a pivotal and trust-by-design part of the FUL life-cycle development, essential for highly regulated and data-sensitive services and applications like healthcare. This article introduces {\veriful}, a reference framework for verifiable FUL that formalizes verification entities, goals, approaches, and metrics. Specifically, we consolidate existing efforts and contribute new insights, concepts, and metrics to this domain. Finally, we highlight research challenges and identify potential applications and developments for verifiable FUL and {\veriful}. This article aims to provide a comprehensive resource for researchers and practitioners to navigate and advance the field of verifiable FUL.
\end{abstract}


\newcommand\mycommfont[1]{\footnotesize\ttfamily\textcolor{blue}{#1}}


\section{Introduction}
\begin{figure*}[t]
	\centering
	\includegraphics[width=\linewidth]{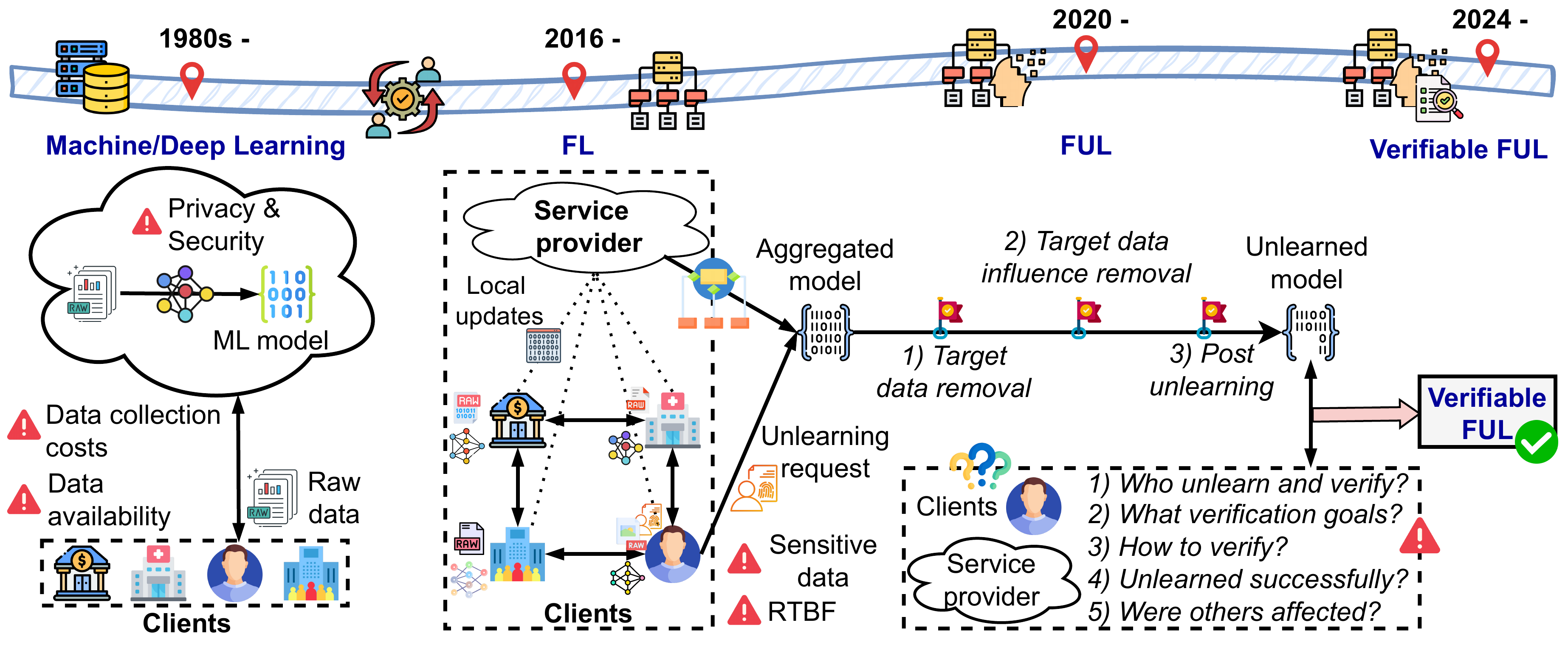}
	\caption{Illustration of the evolution from centralized ML to FL and FUL, highlighting key challenges at each stage and the emerging need for verifiable FUL.}
    \label{fig:evolution_vful}
\end{figure*} 

Federated learning (FL) is a privacy-enhancing collaborative data-sharing and training paradigm in which distributed clients (e.g., end users, edge devices, enterprises, hospitals, or organizations) jointly train a global model under the coordination of a service provider (e.g., a central server/aggregator) while keeping raw data locally, thereby achieving collective intelligence~\cite{mcmahan2017communication}. 
FL has matured from concept to practice, empowering applications from Google keyboard next-word prediction \cite{hard2018federated} to the US cross-center cancer treatments\footnote{https://www.canceralliance.ai/}. Concurrently, data protection regulations, such as the EU's GDPR\footnote{https://gdpr-info.eu/} and California’s CCPA\footnote{https://oag.ca.gov/privacy/ccpa}, have strengthened clients' rights to request removal of their personal data and its influence on trained models, generally called the right to be forgotten (RTBF). For example, hospitals participating in a federated diagnostic network must be able to eliminate a patient's data influence from the global model upon consent revocation. Beyond regulatory compliance, service providers also need to eliminate malicious, noisy, or unlawful data to maintain model integrity and performance. These drivers make the capability of data erasure and its associated influence a first-class requirement in FL systems.

This capability is broadly and technically formalized as machine unlearning (MUL)~\cite{cao2015towards}. 
In MUL, while retraining from scratch offers strong completeness and guarantees that the influence of the target data\footnote{\textit{Target data} refers to the data to be forgotten or the data being requested to unlearn by \textit{data owners/target clients}.} has been eliminated, it incurs prohibitive storage, computational, and time costs, especially in the era of generative AI such as large language models (LLMs). Consequently, research has shifted towards developing more efficient approximate unlearning algorithms that sacrifice the data influence removal completeness for the cost efficiency~\cite{10880482}. This creates \textit{an unlearning-verification gap}, as it is difficult for service providers or clients to verify whether data influence has been removed, and as machine learning (ML) models can compress and retain knowledge from training data~\cite{carlini2019secret}.
%
These challenges are further amplified by the architectural shift towards federated settings \cite{tran2025tofu}. 
Directly adopting MUL methods is nontrivial due to FL-specific constraints, including dynamic client participation, statistical and system heterogeneity among clients, and the service provider’s lack of access to raw data. While these factors have motivated research into federated unlearning (FUL) algorithms, spanning gradient modification and knowledge distillation approaches~\cite{10736348}, underexplored questions persist from clients'
and the service provider's perspectives (see Figure~\ref{fig:evolution_vful}): 
\begin{enumerate}
    \item \textbf{Service provider \& Clients:} \textit{Who will participate in the unlearning and verification process?}
    \item \textbf{Service provider \& Clients:} \textit{What are verification goals needed to be achieved?}
    \item \textbf{Service provider \& Clients:} \textit{How does target data be unlearned?} (accomplished through unlearning algorithms)
    \item \textbf{Target clients:} \textit{How to verify that my data has been unlearned from the trained model?} 
    \item \textbf{Service provider \& Clients:} \textit{Which metrics and evidence are used to evaluate and ensure that target data is being unlearned successfully?}
    \item \textbf{Remaining clients\footnote{This article uses remaining clients and non-target clients interchangeably.}:} \textit{Will my contributions remain intact?}
\end{enumerate}

It is crucial to answer these questions to establish a \textit{trust-by-design} FUL system that upholds the clients' RTBF and their right to \textit{verify} the removal of data influence. Despite progress, a unified framework for verifiable FUL remains lacking. 
Romandini et al. in \cite{10736348} surveyed unlearning algorithms and categorized metrics. Authors in \cite{liu2024survey, jeong2024survey} outlined a workflow and a fine-grained taxonomy specifying who unlearns, who verifies, what is unlearned, and the key lessons and open directions. Gao et al. \cite{gao2024verifi} proposed a mark-to-check protocol that allows the target client to verify the unlearning effect locally, using its hardest and most unique data samples. However, solely proposing unlearning algorithms and reporting metrics does not guarantee faithful unlearning execution. Critical gaps remain, including who conducts verification, what verification goals are across stakeholders (i.e., target clients, non-target clients, and service providers), how to implement verification, how well goals are achieved with selected verification approaches, and how to integrate these elements into a unified framework. If left unaddressed, target clients are often forced to unquestioningly trust the service providers’ and other clients' claims about the unlearning efficacy and integrity, enabling dishonest behaviors (e.g., a service provider appears to have unlearned, passed verification, or fabricated metrics). 
To address this gap, we propose a unified FUL framework, called {\veriful}, which consolidates prior efforts \cite{10736348, liu2024survey, jeong2024survey, gao2024verifi} on identifying who unlearns, who verifies, and how unlearning efficacy is measured, and extends them with a systematic characterization of unlearning verification goals and verification approaches, together with metrics for assessing verification efficacy and efficiency, as well as unlearning efficacy and fidelity, within a coherent framework that we then demonstrate in a healthcare case study. Furthermore, we provide ongoing challenges and future work in this emerging domain.

\section{VeriFUL: VERIFIABLE FUL FRAMEWORK}
\begin{figure*}[t]
	\centering
	\includegraphics[width=\linewidth]{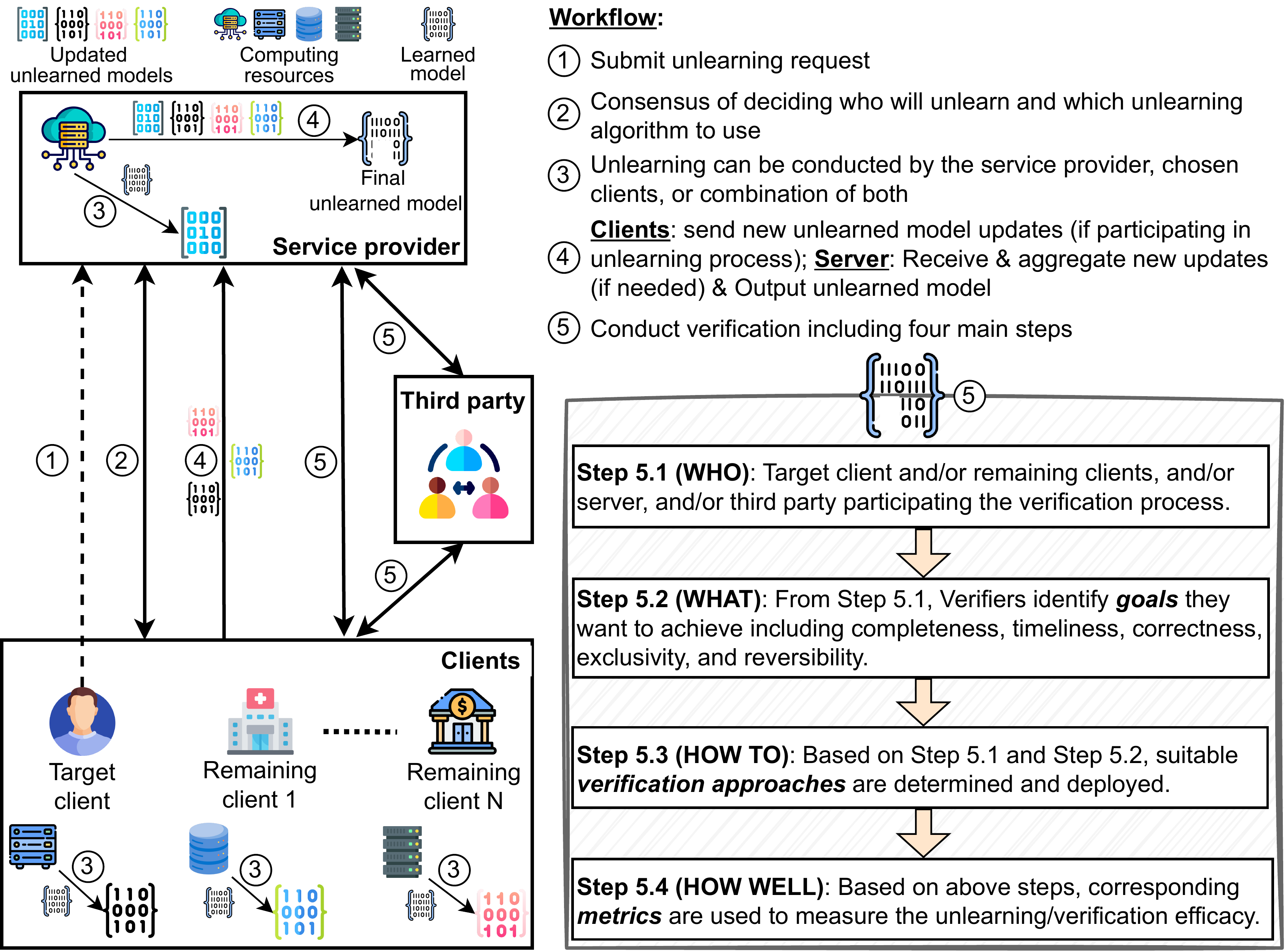}
	\caption{{\veriful} reference framework with workflow - we focus on Step~5 in this article.}
    \label{fig:veriful_framework}
\end{figure*}

This section introduces {\veriful}, specifying WHO, WHAT, HOW TO, and HOW WELL for designing a trust-by-design foundation and verifiable FUL. {\veriful} is a technical control layer that can be built on top of existing FUL systems, helping researchers and practitioners to implement established principles and obligations (e.g., GDPR's RTBF, transparency, and accountability). Figure~\ref{fig:veriful_framework} illustrates components and workflow of {\veriful} with a healthcare case study. Our focus is on the verification stage (Step~5).

\subsection{WHO - Verification Entities}
Verification entities are stakeholders (e.g., clients and service providers) that participate in the verification phase to determine whether the unlearning was faithfully executed (see Step~5.1 in Figure~\ref{fig:veriful_framework}).

\underline{\textbf{Clients}}: comprise the target and remaining clients. Target clients submit unlearning requests and retain \textit{RTBF} and \textit{the right to verify} that their data\footnote{The data can be classified as personal or business ownership-related.} influence has been effectively removed from the global model. Remaining clients verify that their contributions (e.g., updates and data impact) to the global model remain intact. They may act as provers (e.g., proving their updates were correctly computed) and/or as verifiers, assisting checks on target-data removal.

\underline{\textbf{Service providers}}: orchestrate learning and unlearning. When executing unlearning, the service provider must generate and provide verification artifacts such as cryptographic proofs, attestations, audit logs, and evaluation metrics, demonstrating strict adherence to the unlearning algorithm. When unlearning is performed locally by target clients, the service provider may support the secure aggregation of (unlearned) updates and assist in verifying the correctness and exclusivity of the client-side unlearning.

\underline{\textbf{Third parties}}: are independent auditors (e.g., consortium authorities) serving as external validators. They inspect the artifacts supplied by the service provider and/or clients who participate in the verification process, reproduce statistical or cryptographic checks, and certify whether the stated unlearning guarantees hold.

\subsection{WHAT - Verification Goals}
{\veriful} specifies guarantees and expectations a FUL system should meet, including completeness, timeliness, correctness, exclusivity, and reversibility. Clearly defining these goals establishes the foundation for systematic and reliable verification (see Step~5.2).

\underline{\textbf{Completeness}}: ensures that the unlearning process eliminates target data and residual influence from both remaining clients' local models and the final global model. After unlearning, the performance on retained data of the unlearned model should be statistically indistinguishable from that of a retrain-from-scratch model trained on the same retained data, thereby preserving model utility while guaranteeing RTBF. It is noteworthy that {\veriful} mainly focuses on approximate unlearning because exact unlearning via full retraining is intractable in time and costs for modern large-scale models (e.g., LLMs) and in scenarios involving only a small number of unlearning clients or samples. Accordingly, the retrain-from-scratch model is a benchmark to quantify how closely the approximate unlearning algorithm approaches ideal completeness.

\underline{\textbf{Timeliness}}: requires executing unlearning and delivering verification artifacts without undue delay following a valid erasure request. Under the GDPR, unlearning entities must act \textit{without undue delay}, within one month of receiving the request. This deadline can be extended by up to two additional months for particularly complex cases, as long as the target clients are informed (i.e., GDPR Art. 12(3), Art. 17). 

\underline{\textbf{Correctness}}: guarantees that the unlearning algorithm has been carried out exactly as intended, following the prescribed protocol without deviation, omission, or adversarial manipulation. It means that the evidence provided (e.g., cryptographic proofs) must allow target clients and/or auditors to independently check unlearning protocol conformance. Unlike \textit{completeness}, which asks whether all impacts of the target client’s data were removed, correctness asks whether the removal process itself was executed faithfully and auditably.

\begin{table*}[ht!]
  \centering
\small
  \caption{Qualitative Pairwise Interplay of the Five Verification Goals in {\veriful} Framework.}
\label{tab:verification_goals}
\begin{tabular}{|p{0.1\linewidth} |p{0.14\linewidth} |p{0.14\linewidth}|p{0.14\linewidth}|p{0.18\linewidth}|p{0.14\linewidth}|}
\hline
\rowcolor{gray!30}
Goals & Completeness & Timeliness & Correctness & Exclusivity & Reversibility\\
\hline 

{Completeness} & - &  {\downarrowc} latency (heavier compute, storage, and proof generation).  & {\uparrowc} correctness scheme rigor (faithful, auditable, and verifiable execution is enforced). & {\downarrowc} in approximate FUL (aggressive erasure when data features among clients overlap). & {\downarrowc} ease of restore (learned model checkpoints need to be recorded and retrained).  \\
\hline

{Timeliness}  & {\downarrowc} depth forgetting (faster execution may leave residual data influence). & -  & {\downarrowc} guarantee strength (lightweight checks and fewer audits). & {\downarrowc} non-target contribution safeguard (insufficient dependency checks).  & {\downarrowc} rollback fidelity (insufficient recovery paths for unlearned model). \\
\hline

{Correctness} & {\uparrowc} completeness (unlearning algorithm is exactly executed without deviation).  & {\downarrowc} latency (proof and attestation generation and verification add overhead).  & - & {\uparrowc} isolation (detailed and structured logs and non-target data dependency checking).  & {\uparrowc} verifiable rollback (proof-based checkpoints for restoration).\\
\hline

{Exclusivity}  &  {\downarrowc} completeness (isolation of non-target representations constrains forgetting).  & {\downarrowc} speed (safeguard solutions for non-target clients' contributions are integrated). & {\uparrowc} correctness scheme rigor (structured logs, checkpoints, and proof-based validation).  & -   & {\downarrowc} reversibility (restoring data target influence without perturbing others is challenging). \\
\hline

{Reversibility}&  {\downarrowc} completeness (a smooth restoration process limits the forgetting ability). & {\downarrowc} latency (complete rollback demands longer execution time, such as proof generation). & {\uparrowc} correctness (it requires fine-grained traceability and proof-based validation).  & {\downarrowc} exclusivity (reverting the target influence may shift decision boundaries of non-targets).   & -  \\
\hline

\end{tabular}

      \begin{tablenotes}
      \small
    \item \underline{\textbf{Direction}}: Row goal affects Column goal.
    
    \item \underline{\textbf{Legend}}: {\uparrowc} Improves/Strengthens, {\downarrowc} Degrades/Increases costs.

    \end{tablenotes}
\end{table*}

\underline{\textbf{Exclusivity}}: is the assurance that an unlearning procedure operates solely on the data associated with the erasure request, preserving the integrity of all remaining clients’ contributions, particularly in approximate unlearning. From the perspective of remaining clients, it addresses the fundamental concern: \textit{Will my contributions be intact?}. These concerns are reasonable because clients often have overlapping data, and their contributions also relate to their efforts' compensations (e.g., monetary rewards \cite{wang2025unlearning} or a shared learning model) after multiple rounds of participating.

\underline{\textbf{Reversibility}}: ensures that target clients can revoke their unlearning requests, with the system efficiently restoring the forgotten knowledge in a verifiable manner, ensuring performance consistency with the pre-unlearning model.

\noindent \textbf{Discussion:} In {\veriful},  the five verification goals are not separate pillars but interdependent dimensions. Strengthening one goal may impose costs on others and must be explicitly navigated in system design. For example, tighter \textit{completeness} coupled with stronger \textit{correctness} usually demands deep state changes, heavier computation, log retention, and proof generation and verification, slowing down \textit{timeliness}. Retraining from scratch offers the strongest guarantee of eliminating all traces of target data, yet its costs and client availability constraints make it impractical in federated environments. This tension has motivated approximate unlearning methods that enhance \textit{timeliness} while sacrificing some level of completeness \cite{nguyen2024surveymachineunlearning, liu2024survey}. Likewise, mechanisms that enforce strict \textit{exclusivity} raise a huge amount of fine-grained audits and correlated data feature and model representation checks, which require substantial verification and coordinating overhead. Enabling \textit{reversibility} complicates the design by requiring verifiable recovery paths (e.g., weight checkpoints, client selection lists, seeds). Designing deployable systems, therefore, entails negotiating these trade-offs, balancing feasibility, forgetting efficacy,  unlearning fidelity, unlearning efficiency, and the strength of guarantees. Table~\ref{tab:verification_goals} summarizes these correlations.

\subsection{HOW TO - Verification Approaches}
{\veriful} verification approaches specify methods and technologies that provide auditable and verifiable evidence of the unlearning procedure. This is a critical design choice as each offers distinct trade-offs in assurance strength, computational/communication overhead, scalability, and trust assumptions (see Step~5.3 in Figure~\ref{fig:veriful_framework}). 

\underline{\textbf{Cryptographic methods}}: provide strong and mathematically verifiable proof of unlearning adherence. For instance, a zero-knowledge proof (ZKP) \cite{goldwasser2019knowledge} allows a prover (e.g., the service provider or a non-target client) to convince a verifier (e.g., the target client or a trusted third party) that a statement about the unlearning process is true (e.g., the non-target clients' model updates on the non-target datasets were computed correctly using the proposed unlearning algorithm) without revealing the raw data. 

Beyond ZKPs, homomorphic commitments \cite{homo} and verifiable computation systems \cite{ver} can also strengthen unlearning guarantees in federated settings. Homomorphic commitments allow for a verifiable proof that the forgotten update has been correctly subtracted from the aggregation.
Similarly, verifiable computation systems such as SNARKs allow the service provider to prove that it executed the prescribed unlearning algorithm correctly when recomputing the global parameters, so that clients can verify adherence to the protocol without re-running expensive computations.

The strength of cryptographic methods is their ability to provide strong, objective guarantees for \textit{exclusivity} and \textit{correctness}, which is a prerequisite for \textit{completeness} assessments. However, a key challenge is their significant proof size, communication overhead, circuit/specification engineering, and latency impacts on \textit{timeliness}, especially for large AI models \cite{10966041}.

\underline{\textbf{Hardware-based attestation}}: 
leverages a trusted execution environment (TEE) such as Intel Software Guard Extensions \cite{mckeen2013innovative} to perform unlearning and aggregation within an isolated enclave, providing integrity and confidentiality guarantees to the unlearning code and data even if the host system is compromised. A TEE can generate a remote attestation, a cryptographically signed report proving to a verifier that the expected unlearning code has been loaded into a genuine enclave. This contributes to \textit{correctness} by assuring that the unlearning process is initiated with the intended code, though it does not guarantee that the code executes to completion without runtime interference or side-channel leakage. Compared to pure cryptographic proofs, TEEs achieve this with much lower performance overhead, thereby supporting \textit{timeliness}. 
Most current TEEs face strict enclave memory limits, which make them unsuitable for storing and processing large AI models. As a result, their use in FUL is often restricted to verifying smaller model fragments or coordinating the aggregation process.

\begin{table*}[ht!]
  \centering
  \small
  \caption{Comparative Analysis of Verification Approaches in {\veriful} Framework}
\begin{tabular}{|>{\columncolor[gray]{0.85}}p{1.7cm}|p{2.4cm}|p{4cm}|p{4cm}|p{4cm}|}
    \hline
    \rowcolor{gray!30}
    \textbf{Approach} & \textbf{Overhead per unlearning request} & \textbf{Scalability w.r.t model size (M) and no. of clients (N)} & \textbf{Security strength} & \textbf{Use in }{\veriful}\\
    \hline
    {\multirow [c]{2}{*}{\parbox{1.5cm}{Cryptographic proofs}}} & \textbf{High}. Large proof size and proof generation time. &  \textbf{Low}. Degrades when M and N grow.  &  \textbf{High}. Strong, formal guarantees for \textit{correctness} and \textit{exclusivity}. &  Unlearning in highly-regulated and adversarial settings with moderate M, N, and the number of unlearning requests.  \\
    \hline
    {\multirow [c]{3}{*}{\parbox{1.5cm}{Hardware-based attestation}}}  &  \textbf{Low-moderate}. Near-native execution speed and costs associated with attestation. & \textbf{Low-moderate}. Limited by enclave memory limits; scales well with N but is unsuitable for large M [2].  &  \textbf{Moderate}. Strong guarantees for \textit{correctness} (e.g., executing unlearning code and model updates/aggregation) inside the enclave, assuming the trusted hardware; vulnerable to side-channel attacks. & Ensure \textit{correctness} and \textit{timeliness} for time-critical applications with small/moderate M; execute sensitive unlearning parts.   \\
    \hline
    {\multirow [c]{2}{*}{\parbox{1.5cm}{DLT-based auditing}}} &  \textbf{High}. Significant storage and latency due to consensus and transaction history.  &  \textbf{Moderate}. Independent of M with off-chain storage; handle large N; ledger storage and transaction confirmation latency grow with the number of unlearning events.  & \textbf{Moderate}. Strong non-repudiation and tamper-evident logging for \textit{reversibility}; lacks guarantees for \textit{correctness}. &  Regulated consortia requiring auditable history, accountability, and controlled rollback under moderate unlearning intensity. \\
    \hline
    {\multirow [c]{2}{*}{\parbox{1.5cm}{Active testing}}}  & \textbf{Low}. Per-verifier overhead (i.e., unlearned model evaluation and statistical/attack analysis).  &  \textbf{High}. Scales well with M and N, but requires access to the final unlearned model. &  \textbf{Low}. Empirical evidence towards \textit{completeness} and \textit{exclusivity}, but cannot guarantee \textit{correctness} itself.  & Periodic monitoring of the unlearning cycle, serving as a mandatory and lightweight complement to verification methods.  \\
    \hline

  \end{tabular}
  \label{tab:verification_approaches}
\end{table*}

\underline{\textbf{Distributed ledger technology-based auditing:}} leverages distributed ledger technologies (DLTs), such as blockchain, to create a transparent, chronological, and immutable audit trail for the unlearning workflow. In {\veriful}, a permissioned blockchain (e.g., Hyperledger Fabric v2.x\footnote{https://hyperledger-fabric.readthedocs.io/en/release-2.5/}), which offers higher throughput, lower latency, and fine-grained access control than public blockchains \cite{nguyen2025blockchain}, is jointly operated by validator nodes (e.g., one or two anchor clients and a service provider), recording signed transactions that capture unlearning requests, role assignments, chosen learning/unlearning algorithms, key hyperparameters, model learning/unlearning checkpoints and their cryptographic commitments (e.g., hashes of model updates), as well as verification outcomes or disputes. Besides, the remaining clients that perform the unlearning act as lightweight clients. These clients submit signed transactions and can verify the audit trail without participating in the heavy consensus. Rather than storing model updates on-chain, which is technically and economically costly, the blockchain only records cryptographic hashes of model updates and key artifacts, while the corresponding encrypted artifacts remain in off-chain storage \cite{nguyen2025blockchain}. During an audit or rollback, these hashes allow verifiers to check that a recovered model state is consistent with the recorded unlearning actions. Consensus mechanisms such as Practical Byzantine Fault Tolerance \cite{10937129} and its variants, which are scalable with the number of unlearning requests in terms of low latency and high throughput compared to proof of work and proof of stake, and offer deterministic finality, can be an appropriate choice for DLT-based unlearning auditing. In summary, DLT-based auditing primarily strengthens \textit{correctness} and \textit{reversibility}, while addressing \textit{timeliness} and integrating ZKPs for mathematical \textit{correctness} of unlearning execution remain important directions. 

\underline{\textbf{Active testing}}: 
Distinct from DLT-based auditing, which provides immutable formal logs, active testing refers to empirical probing and assessment of the released unlearned model through evaluation and analysis. Instead of providing proof of correct execution like the approaches above, it reinforces trust by validating and providing the outcomes against targeted checks. For example, the verifier can test the unlearned model on chosen datasets (e.g., target client's data), analyze differences in model updates before and after unlearning, or repurpose security and privacy attacks as audit tools (e.g., membership inference attacks-MIAs). This approach offers flexible, intuitive, and measurable results. Solely using this approach helps us to manage \textit{timeliness} due to its low computational overhead, but it cannot, by itself, verify \textit{correctness}, as a malicious prover may fabricate the results without actually running the unlearning algorithm.

\noindent\textbf{Discussion:} We analyze and summarize overhead, scalability, security, and the use in {\veriful} of verification approaches in Table~\ref{tab:verification_approaches}. 

\subsection{HOW WELL - Verification Metrics}

Verification metrics quantify how well verification goals are met under a given approach. Existing surveys summarize candidates \cite{10736348, 10880482, liu2024survey, jeong2024survey} but lack a unified standard for definitions, evaluation procedures, and acceptance criteria. Building on these efforts, {\veriful} consolidates, systematizes, and extends the metric space into a coherent taxonomy (see Table~\ref{tab:verification_matrics} and Step~5.4 in Figure~\ref{fig:veriful_framework}).

\subsection{A Healthcare Case Study} To ground {\veriful} in a practical setting, we consider a healthcare consortium where multiple hospitals and cancer centers collaboratively train a global AI model for cancer treatment (e.g., the Cancer AI Alliance in the US\footnote{https://www.canceralliance.ai/}) without sharing raw patient data. An aggregation service, operated by the consortium or a service provider, coordinates training and performs model aggregation. 

Suppose a participating hospital (i.e., the target hospital) needs to delete a subset of patient records and remove their influence on the global cancer treatment model. 
The target hospital submits an unlearning request to the consortium (Step~1). The consortium executes a consensus procedure to determine: (i) which entities will perform the unlearning (e.g., the aggregation service, the target hospital, and/or non-target hospitals) and (ii) which unlearning algorithm will be executed (Step~2). For example, an approximate unlearning algorithm such as FUSED \cite{zhong2025unlearning}, which identifies a few knowledge-sensitive model layers, attaches sparse trainable adapters, freezes the base model, and retrains only these adapters on the retained data, can be deployed across the aggregation service, the target hospital, and non-target hospitals to produce unlearned model updates, so that retained knowledge overwrites forgotten knowledge.
Unlike standard FUL, {\veriful} ensures trust by embedding verification artifacts directly into the unlearning lifecycle.
The selected entities proceed to perform the unlearning procedure (Step~3), and the aggregation service then aggregates these unlearned model updates. They generate verifiable evidence, including cryptographic commitments to model updates, training data, hyperparameters, and ZKP-based proofs of correct execution.

Therefore, {\veriful} enables a robust verification stage as shown in Step~5. 
Here, the participating entities agree on the set of verifiers, potentially which may include the target hospital, non-target hospitals, the aggregation service, and/or independent third-party auditors (Step~5.1). These verifiers with the target hospital in agreement then define the specific verification goals (e.g., completeness, correctness), tailored to the sensitivity of the data and consortium policies (Step~5.2). Based on these goals, they utilize the pre-generated evidence to validate the unlearning process and conduct active testing (e.g., testing the unlearned cancer treatment model performance) (Step~5.3).
Finally, quantitative verification metrics (e.g., performance delta of the unlearned cancer treatment model, and PVSR) are computed and analyzed to assess how well each goal has been met, thereby providing measurable assurance to the target hospital and other stakeholders regarding the forgetting efficacy, efficiency, fidelity, and integrity of the unlearning (Step~5.4). We integrate these steps and configurations as a request-level protocol around the unlearning workflow. This makes explicit how {\veriful} can guide the design of deployable verifiable FUL pipelines, while remaining agnostic to the particular unlearning algorithms and verification primitives, thereby accommodating diverse infrastructure, application requirements, and regulatory constraints.

\begin{table*}[ht!]
  \centering
  \small
  \caption{Taxonomy of {\veriful} Verification Metrics.}
  \begin{tabular}{|p{0.5cm}|p{2.4cm}|p{4.2cm}|p{5.2cm}|p{3cm}|}
    \hline
    \rowcolor{gray!30}
    \textbf{Goal} & \textbf{Metric} & \textbf{What it measures} & \textbf{Typical operationalization} & \textbf{Acceptance criteria} \\
    \hline
   {\multirow [c]{11}{*}{\rotatebox{90}{\textbf{Completeness}}}} & \textit{Performance delta} &  Change in learned model performance before vs. after unlearning.  &  Evaluate accuracy/loss/F1 score of unlearned model on the target dataset.  & Significant drop (e.g., $>90\%$) \\ \cline{2-5}

     & \textit{Residual-influence distance} &  Divergence between unlearned model and a retrained-from-scratch model.  & Compute KL/Wasserstein distance between model distributions or parameters on retained data.  &  Distance below predefined threshold. \\\cline{2-5}

    {} & \textit{Probe success rate} &  Vulnerability to adversarial probes testing residual target data memorization.  &  Measures backdoor ASR on triggered patterns, or MIA accuracy and influence-function scores on target data. &  Low ASR/accuracy/influence score.\\\hline

    {\multirow [c]{6}{*}{\rotatebox{90}{\textbf{Timeliness}}}}  &  \textit{Latency} & Total time from request receipt to verified model availability. &  Measure end-to-end time: consensus, unlearning execution, aggregation, verification (Step~2-Step~5, see Figure~\ref{fig:veriful_framework}). & Meets SLA/deadline (e.g., $<24$ hours and $<1$ month for GDPR). \\\cline{2-5}

    {} & \textit{Throughput} &  Number of unlearning requests completed per unit time.  &  Aggregated logs of completed requests (e.g., per second/minute/hour/day). & Throughput $\geq$ pre-defined unlearning target rate \\\hline

    {\multirow [c]{7}{*}{\rotatebox{90}{\textbf{Correctness}}}}  &  \textit{Proof verification success rate (PVSR)} & Proportion of valid proofs accepted by verifiers.  &  Generate cryptographic proofs for each unlearning step; verifiers validate. & PVSR = 1.0 for all proofs. Any failure triggers audit. \\\cline{2-5}

    {} & \textit{Auditing score} &  Quantitative score from third-party audit.  & Auditor examines logs, checkpoints, and assigns score (e.g., 0–100). & Score above audit threshold (e.g., $\geq$ 90/100).\\\hline

    {\multirow [c]{9}{*}{\rotatebox{90}{\textbf{Exclusivity}}}}  &  \textit{Performance-level stability} & Change in performance on remaining clients’ local test sets. & Evaluate unlearned model on each remaining client’s fixed test set before/after unlearning. & Performance change $\leq$ small delta (e.g., $\pm 0.5\%$ accuracy). \\\cline{2-5}

    {} & \textit{Parameter-level stability} &  Similarity of a remaining client’s model updates before/after unlearning.  & May compute cosine similarity between client’s model updates. & High similarity value (e.g., $\geq 0.95$ means highly intact). \\\cline{2-5}

    {} & \textit{Behaviour-level stability} &  Divergence in model output distributions on remaining client data before/after unlearning.  &  May compute Wasserstein distance between output distributions. & Distance value $\approx 0$ means highly intact. \\\hline
    
    {\multirow [c]{9}{*}{\rotatebox{90}{\textbf{Reversibility}}}}  & \textit{Performance consistency} & Performance difference between restored and pre-unlearning model. &  Evaluate both models on same test set. & Change $\leq$ threshold (e.g., $<0.5$\% accuracy). \\\cline{2-5}

    {} & \textit{Restoration latency }&  Time to revert unlearned model to its pre-unlearning state.  &  Measure total time from revert request to restored model availability. & Latency $\leq$ latency SLO (e.g., 30 minutes).\\\cline{2-5}

    {} & \textit{Model state integrity} &  Cryptographic comparison of model states pre/post-restoration. &  Compare hashes of model checkpoints (under deterministic training). & Exact hash match (if feasible).
 \\\hline
  \end{tabular}
    
    \begin{tablenotes}
      \scriptsize
    \item *** \textit{KL, SLA, and SLO mean Kullback-Leibler divergence, Service Level Agreement, and Service Level Objective, respectively}.
    
    \item *** \textit{Setting of pre-defined thresholds, SLAs, and SLOs are dependent on application/service/existing infrastructure/client needs}.
    \end{tablenotes}
  \label{tab:verification_matrics}
\end{table*}
\color{black}
\section{CHALLENGES}
Given the realization of verifiable FUL remaining in its infancy, we need to tackle several challenges.

\subsection{The Co-Evolution of Unlearning Algorithms and Verifiability}
The enforcement of RTBF, the need to verify, and the accountability of service providers mandate that unlearning algorithms be designed to be inherently auditable and verifiable, rather than treating verification as a post-hoc addition. This shifts the algorithm design from the sole objective of approximating the trained model to a dual goal of both achieving unlearning efficacy/efficiency/fidelity while ensuring compatibility with verification approaches. This challenge is particularly for approximate unlearning.
Unlike full retraining, verifying an approximation requires proving that the unlearned model operates within an acceptable/pre-defined threshold of residual data influence, which is ambiguous to define and audit. Consequently, the unlearning workflow must integrate detailed model transition stage logs, model checkpoints, and proof and commitment generation schemes to satisfy verification goals, such as \textit{correctness} and \textit{reversibility}, thereby reducing client concerns and promoting trust. However, integrating these proof-generating and model-state management into the unlearning workflow may incur substantial latency and overhead.

\subsection{Efficiency vs. Privacy Compliance}
Strict enforcement of RTBF in federated settings can conflict with the need for efficient model training and deployment. When target data is highly informative or unlearning requests occur at scale, the removal and verification processes may substantially degrade the model’s generalization ability, affecting both convergence and downstream task performance. Furthermore, the additional overhead from unlearning and verification, {especially using cryptographic proofs or DLTs, in terms of computation and communication}, delays the release of stable global models. These challenges are further magnified in resource-constrained environments where clients have limited computing, memory, and storage to support repeated and concurrent unlearning and verifiability operations.

\subsection{Privacy and Security}
In the current landscape, a client’s intent to unlearn becomes explicitly disclosed through direct requests, which exposes the request to the service provider and other clients. While necessary for coordination and auditability, this transparency opens the door for adversarial behaviors. Particularly, the service provider may selectively omit, delay, or ignore unlearning requests, thereby undermining both the confidentiality and enforceability of the unlearning process. 
Moreover, using \textit{active testing-based verification} (e.g., backdoor auditing and MIAs) may introduce unintended vulnerabilities \cite{gao2024verifi}. Malicious verifiers (e.g., the service providers, non-target clients) could exploit these techniques to extract sensitive information, infer training set membership, or tamper with model behavior. Therefore, designing verification mechanisms with robustness against adversarial misuse remains a pressing challenge.

\subsection{Incentive mechanisms}
Designing incentive mechanisms for verifiable FUL remains a significant challenge. Clients may exhibit strategic or inconsistent behaviors, causing a dynamic environment and carryover effects, such as submitting unlearning requests, revoking them later, participating selectively in verification phases, or engaging in learning only when it directly benefits them. These behaviors introduce significant inefficiencies to the service provider and other clients, including wasted computational resources, delayed model updates, and increased coordination overhead. Furthermore, the costs of unlearning and verification are often unevenly distributed among participants and private to the service provider, creating information asymmetry among them, complicating the incentive mechanism design. 

\subsection{Scalability}
Scalability is a fundamental challenge for practical verifiable FUL systems, which must operate across massive, heterogeneous datasets and accommodate dynamic and resource-constrained clients that may be unable to support intensive unlearning and verification tasks. The computational and communication costs of verification methods, such as the cryptographic proofs required for strong \textit{correctness} guarantees, become prohibitive when applied to the large-scale models, including LLMs with billions of parameters. A related challenge is managing the throughput and latency occurring when a high volume of concurrent unlearning requests happens, which remains an open question. These scalability bottlenecks are a critical design constraint that directly affects deployment feasibility, as they can delay the release of stable global models and risk pushing the system toward an unlearning saturation threshold, where the cumulative impact of unlearning may lead to catastrophic or irreversible utility loss, severely affecting the model generalization and \textit{exclusivity}.

\section{THE ROAD AHEAD}
Building on {\veriful}, we now identify four promising research directions for future investigation, including the design of \textit{application-specific solutions}, a shift towards \textit{target client-executed unlearning}, the development of \textit{unlearning-driven incentive and verification mechanisms}, and the formulation of \textit{predictive scaling laws for verifiable FUL}. We highlight each of these directions below.

\subsection{Application-specific Solutions}
A one-size-fits-all {\veriful} is infeasible. Solutions should be tailored to the specific application requirement. In highly regulated, privacy-sensitive sectors such as healthcare, an unlearning request is often driven by a client's exercise of their RTBF. The system must prioritize provable \textit{completeness} to ensure the thorough removal of data influence and strict \textit{correctness} to provide an auditable trail for regulatory compliance, accepting weaker \textit{timeliness} if necessary. When unlearning is service provider-initiated to remove illegal or malicious data impacts, the primary objective is to maintain model integrity and inference performance.
Verification, therefore, focuses on \textit{completeness} to remove the harmful impact and \textit{exclusivity} to protect remaining clients' contributions, alongside empirical metrics that validate the model's inference post-unlearning. Some applications require \textit{reversibility} to accommodate clients' revoked requests
(e.g., opt-in/opt-out on an online federated education platform with federated personalization). Ultimately, designing a deployable system requires negotiating the trade-offs between the verification goals outlined in the {\veriful} framework to align with each application's specific legal, ethical, and operational context, motivating standardized benchmarks that capture these differing objective profiles.

\subsection{Target Client-Executed Unlearning} 
While current verifiable FUL research typically assumes that unlearning is performed by the service provider and/or remaining clients \cite{liu2024survey}, we envision an emerging direction that shifts the unlearning computation to the target clients. This client-centric approach alters the trust model for verification by reducing dependence on the service provider and other clients, while providing target clients greater control over unlearning processes. However, it introduces challenges in verifying that the target client's local unlearning computation was performed \textit{correctly} and \textit{exclusively} on the target data. Malicious clients may exploit this process to inject poisoned updates or tamper with the global model under the guise of unlearning, posing risks to both the service provider and other clients. Moreover, FUL clients may be resource-constrained, which hinders their local unlearning and verification execution. Therefore, it is necessary to develop lightweight and scalable verification protocols for client-side unlearning.

\subsection{Incentive-guided Unlearning and Verification}
Engaging in the unlearning and verification processes imposes additional communication and computation overheads and privacy leakage risks on participants. Without fair and transparent incentives, a verifiable FUL system is likely unsustainable. Frequent unlearning requests with high intensity (e.g., removal of large data volumes driven by strict privacy preferences) may cause irreversible degradation in model performance. Future research should explore incentive mechanisms that reward participants in proportion to their verifiably measured resource contributions while ensuring system-wide fairness. Particularly, designing game-theoretic-based incentive mechanisms can enable service providers and target clients to negotiate acceptable trade-offs between privacy preferences, performance degradation, and economic rewards. Such mechanisms sustain participation in training, unlearning, and verification.
For instance, a transparent incentive layer atop {\veriful} framework can be implemented using blockchain. Client contributions to unlearning and verification are recorded on-chain, enabling automated auditing and reward allocation via smart contracts.

\subsection{Predictive Scaling Laws for Verifiable FUL}
An open direction is an empirical framework to quantify the cumulative impact of unlearning on global model utility (e.g., performance, verification efficacy). Applying scaling-law analysis can characterize how model utility varies with unlearning intensity (e.g., volume, type, distribution of target data, and request dynamics). Then, saturation points are identified, which further forgetting leads to catastrophic utility loss or renders verification metrics unreliable. 
These insights enable service providers and clients to make informed trade-offs between RTBF and preservation of model utility.

\section{CONCLUSION}
In this article, we have introduced verifiable FUL ({\veriful}), a new paradigm in which verification is integrated by design to uphold RTBF and the right to verify, thereby strengthening client trust and control of privacy. We have proposed a reference verifiable FUL framework {\veriful}, detailing the verification entities, multifaceted goals, technical verification approaches, and quantitative metrics. We have presented the key challenges to enable the practical deployment of verifiable FUL. Finally, we have highlighted promising research directions to guide future advancements in this domain.



\section*{Acknowledgments}
This research has been conducted with financial support of Taighde Éireann – Research Ireland under Grant number 18/CRT/6222; the School of Computer Science and Statistics, Trinity College Dublin; and the ENSURE-6G project, which is funded by the MSCA under grant agreement No. 101182933. The work has also been supported in part by Research Ireland under the European Innovation Council CHIST-ERA SHIELD project (Project No. 216449, Award No. 19226). For the purpose of Open Access, the author has applied a CC BY public copyright licence to any Author Accepted Manuscript version arising from this submission.


\bibliographystyle{IEEEtran}
\bibliography{Refs}

\end{document}